\documentclass[10pt,journal,twoside]{IEEEtran}
\usepackage{enumerate}
 \usepackage{amssymb}
\usepackage[dvips]{graphicx}
\usepackage[cmex10]{amsmath}
\usepackage{amsthm}
\usepackage{latexsym,bm}
\usepackage{color}
\usepackage{cite}
\usepackage{subfigure}
\usepackage{multirow}
\usepackage{longtable}
\usepackage{diagbox}[2011/11/22]
\usepackage{cases}
\usepackage{framed}
\begin{document}

\title{Rate-Diverse Gaussian Multiple Access: Efficient Encoder and Decoder Designs
 }
\author{  \IEEEauthorblockN{Pingping Chen, \emph{Member, IEEE}, Long Shi, \emph{Member, IEEE}, Yi Fang, \emph{Member, IEEE}, \\
Francis C. M. Lau,  \emph{Senior Member, IEEE}, and Jun Cheng,  \emph{Member, IEEE} }  \vspace{-0.4cm}
\thanks{ P. Chen is with the School of Physics and Information Engineering,
Fuzhou University, Fujian 350002, China (e-mail:  ppchen.xm@gmail.com).
}
\thanks{L. Shi is with the Department of Electronic and Optical Engineering, Nanjing University of Science and Technology, Nanjing 210094, China (slong1007@gmail.com).}
\thanks{
Y. Fang is with the School of Information Engineering, Guangdong University
of Technology, Guangzhou 510006, China (e-mail: fangyi@gdut.edu.cn).
}
\thanks{
F. C. M. Lau is with the Department of Electronic and Information Engineering, Hong Kong Polytechnic University, Kowloon, Hong Kong (e-mail:
francis\-cm.lau@polyu.edu.hk). }
\thanks{Jun Cheng is with the Department of Intelligent Information Engineering and Science, Doshisha University, Kyoto 610-0321, Japan (e-mail:
jcheng@ieee.org).}
}
 \maketitle
\begin{abstract}
In this work, we develop a pair of rate-diverse encoder and decoder   for a two-user Gaussian multiple access channel (GMAC). The proposed scheme enables the users to transmit with the same codeword length but different coding rates under diverse user channel conditions.
First, we propose the row-combining (RC) method
and row-extending (RE) method  to design  practical low-density parity-check (LDPC) channel
codes for rate-diverse GMAC.
 Second, we develop an
iterative rate-diverse joint user messages decoding (RDJD) algorithm for GMAC, where all user messages are decoded with a single parity-check
matrix. In contrast to the conventional network-coded multiple access (NCMA) and compute-forward multiple access (CFMA) schemes  that first recover a linear
combination of the transmitted codewords and then decode both user messages, this work can decode both the user messages simultaneously.  Extrinsic information transfer (EXIT) chart
analysis and simulation results indicate that RDJD  can achieve
  gains up to 1.0 dB over NCMA and CFMA in the two-user GMAC. In particular, we show that there exists an optimal rate allocation for the two users to achieve the best decoding performance given the
channel conditions and  sum rate.

\end{abstract}


\vspace{-0.2cm}

\section{Introduction}
{
An interference channel  has been considered as a
canonical model to understand the design principles of
cellular networks with inter-cell interference \cite{han1}.
A Gaussian multiple access channel (GMAC) is a simple interference channel where
 two users communicate with one receiver in
the presence of additive white Gaussian noise (AWGN). The
capacity region of the two-user GMAC has been
characterized in \cite{sic1,sic2}, where the corner points of the capacity region can
be achieved via single user decoding, and  the points in between can be achieved by time sharing. It is also shown that
any rate pair in the capacity region can be attained by
rate splitting or joint decoding without time sharing \cite{han5,han6}. Some enabling solutions of GMAC have been extensively studied in the field of coding and information theory, such as superposition coding \cite{added6,SICNEW3}, successive interference
cancellation (SIC), and the message passing algorithm (MPA)  \cite{ccc3,dddd3}.
As the best known achievable strategy,
Han-Kobayashi scheme  is  a joint unique decoding of messages for the two-user interference channel  \cite{han2}. Each user splits the information into private and common parts by superposition encoding. Both
parts are optimally recovered by using simultaneous
joint decoding \cite{han3,han4}.  Since the recovery of the
   private part can be simply  decoded as in the point-to-point communications, the main challenge lies in the transmission and recovery of the common messages, where each receiver treats the
interference channel as a MAC.


For a GMAC, multiple users transmit their messages simultaneously to a receiver. 
  Among the various types of receivers, SIC has been widely studied because of
its simplicity, where serial decision and cancellation are performed \cite{bb6, ccc12, SICNEW1,SICNEW2}.
In a SIC-aided GMAC,
the received signal powers from different users at the receiver are significantly different. Thus, the receiver first
decodes the message corresponding to the strongest signal and removes this message from its observation.
 The corner points of the two-user capacity  can achieved by SIC  in  GMAC, while other points can be achieved by
time-sharing, rate-splitting, or joint decoding. In particular, capacity-approaching low-density parity-check (LDPC) codes have been  designed
for two-user GMAC by using joint decoders at the receiver \cite{RMAC0, RMAC1, RMAC2}.  A more recent study has optimized the power ratio of the two users to minimize the required SNR for a given rate under the joint decoding in a GMAC \cite{RMAC3}.

Not until very recently has the notion of  physical-layer
network coding (PNC) been  adopted into the two-user GMAC, referred to as network-coded
multiple-access (NCMA) \cite{ccc6}.
The core of PNC is to turn  mutual interference between signals received from
simultaneously transmitting users into useful network-coded (NC)
messages \cite{bib555,you1}. As a prevailing decoding in PNC, XOR-based channel decoding
(XOR-CD) aims to decode bitwise NC messages of user messages.
The first work on NCMA has used XOR-CD and multiuser
decoding (MUD)  to first  decode the
NC message and  then both user messages \cite{ccc6}.
Later, to achieve a high-efficient transmission,   a NCMA system that operates with
high-order modulations of QPSK and 16-QAM has been developed\cite{you2}.
  To exploit
different channel conditions,
 different modulations for distinct users  have been adopted to realize modulation-diverse NCMA with the same coding rate\cite{nonp}. It was shown that the
system throughput of NCMA with BPSK+QPSK outperforms those of modulation-homogeneous NCMA.
 More recently,    a NCMA downlink system for  unmanned aerial vehicle (UAV) communications has been developed to
improve system throughput  by NCMA-based superposition
coding\cite{uman}.

In another line of PNC-based GMAC, compute-and-forward multiple access (CFMA)
generalizes PNC to multi-user networks
by utilizing structured nested lattice codes \cite{CCFA,TAO}.
The authors have extended compute-and-forward
to a GMAC  from a theoretical perspective  \cite{ITCFMA}. By appropriately scaling
the lattice codes,  it can attain the dominant face
of the capacity region of GMAC
 with a single-user decoder without time sharing.   Following this study,
practical off-the-shelf LDPC codes and  efficient decoding algorithms have been adopted for CFMA\cite{CFMA}.  Similar to NCMA, CFMA first decodes a linear NC combination of two codewords by XOR-CD, and then recovers the individual user codewords.
 It has been shown that   CFMA can  achieve some rate pairs on the dominant face of the capacity region, whereas the conventional
SIC fails to attain these rate pairs. However,  the achievable rate of XOR-CD in a two-user GMAC is limited by the worse user channel even when the other channel is error-free. In this context,  CFMA may not perform well over the user channels that are significantly diverse. This is due to the fact that CFMA   decodes the NC message by using the parity-check matrix of the better user channel only.

%

%
%

 In this paper, we propose a two-user uplink rate-diverse GMAC  over diverse channel conditions. We call the user  with the better channel condition  as the strong user and that with the worse channel as the weak user. The proposed system allows the users to adopt different channel codes, i.e.,   a low-rate code  for the weak user and  a high-rate code   for the strong user. This work is primarily motivated by the observation that the achievable rate of XOR-CD in a two-user GMAC is limited by the worse user channel even though the other channel is error-free, and this limitation prevents the strong user from fully exploiting its superior channel condition. In this context, the goal of this paper lies in the encoder and decoder designs that can fully exploit the diverse channel conditions in a two-user GMAC.
    Our main contributions are as follows.

\begin{enumerate}

    \item For the encoder of the proposed scheme, we design practical LDPC codes for different users  by two methods, i.e., row-combining (RC) method and row-extending (RE) method. The core of these methods is  that the two channel codes share a common parity-check matrix with a high code rate. In particular, the parity-check matrix of the low-rate code owns a residual matrix alongside with the common matrix, and the parity-check matrix of the high-rate code is embedded in that of the low-rate code.

\item For the decoder of the proposed scheme, we propose an iterative rate-diverse joint user-message decoding (RDJD) algorithm to decode user messages within the  low-rate parity-check matrix.  The proposed decoder consists of a joint user channel decoder (JUD)  to decode the joint user message and  a residual user decoder (RUD) to decode the individual user messages. It differs from   NCMA and CFMA which first decode  the NC message and then recover both user messages based on the decoded NC messages.

\item  We investigate the  error   performance of the iterative
RDJD using an extrinsic information transfer (EXIT) chart.
 Analytical results show that iterative decoding between JUD and RUD provides large gains over the non-iterative one. Our simulations further demonstrate that iterative RDJD achieves  significant performance gains   over NCMA and CFMA. This gain can be up to 1 dB when the user channels are largely  different. In particular, both analytical and simulation results indicate  that there exists an optimal rate pair for the two users to achieve the best performance under given channel conditions. Moreover, we show that the proposed RDJD can approach  the rate pair on the dominant
face of the capacity region without rate-splitting or time-sharing.


%

%
%
\end{enumerate}

The remainder of this paper is organized as follows. Section II presents the related background for   GMAC. Section III puts forth the concept of rate-diverse GMAC and elaborates its encoder and decoder.
 Section IV investigates the iterative decoding of the proposed scheme by means of an EXIT chart analysis. Section V presents the performance of the rate-diverse GMAC and validates its advantages. Finally, Section VI concludes the paper.

%
}

{
\section{Related background}

%
%
%
%
%
%
%
%
%

%
%
%
 In this paper, we use a boldface letter to denote a vector and the corresponding italic letter to denote a symbol within the vector. For example, $\textbf{x}$ is a vector, and $x$ is a symbol within the vector.

 \subsection{System Model
}
  We consider  the two-user GMAC  in Fig.~\ref{fig:LONG}. Let  $\mathbb{F}_q$ denote a $q$-ary finite field   and $q$ is a prime power. Let $\mathbf{u}_1$  and $\mathbf{u}_2$  be two source message vectors of
  lengths $K_1$ and $K_2$ for users 1 and 2, respectively, $u_1,u_2\in \mathbb{F}_q$.   Channel codes are used to encode $\mathbf{u}_1$ and $\mathbf{u}_2$  into two length-$L$ vectors
  $\mathbf{c}_1$ and $\mathbf{c}_2$ with rates of $R_1=K_1/L$ and  $R_2=K_2/L$, respectively, where $c_1,c_2\in  \mathbb{F}_q$.  The channel codes with $R_1$  and $ R_2$ have parity-check matrices $\mathbf{H}_1$ and $\mathbf{H}_2$, respectively.  We use $\Gamma_1$ and $\Gamma_2$ to denote the channel coding for  the  two users, i.e.,
\begin{equation}
 \mathbf{c}_1=\Gamma_1(\mathbf{u}_1), \mathbf{c}_2=\Gamma_2(\mathbf{u}_2).
 \label{eq:eq1}
 \end{equation}
 Users 1 and 2 generate two signal vectors $\mathbf{x}_1$ and $\mathbf{x}_2$ by symbol-wise modulations, i.e., $x_1=\varphi_1(c_1)$ and $x_2=\varphi_2(c_2)$,
 where $\varphi_m$ defines the modulation from $c_m$ to $x_m$ of user $m$. Let $\chi_m$ denote the constellation set   of user $m$. The cardinality of $\chi_m$ is $q$, i.e., $|\chi_m|=q$  (by cardinality, we mean the number of distinct elements
in the set), and    $x_m\in\chi_m$, $m=1,2$.

 \begin{figure}[!t]
\center
\includegraphics[width=3.3in,height=1.0in]{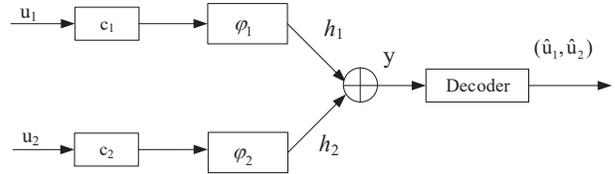}
\vspace{-0.3cm}
\caption{GMAC channel model.
}
\label{fig:LONG}
\end{figure}


Let $h_1$ and $h_2$  denote the channel coefficients from user  1 and user 2 to the receiver, respectively. Assuming perfect
synchronization between the two users, the  signal at the receiver is given by
\begin{equation} \mathbf{y} = h_1\mathbf{x}_1 + h_2\mathbf{x}_2 + \mathbf{z}  \label{eq:eq2},  \end{equation}
where the vector $\mathbf{z}$  contains white Gaussian noise samples with zero mean and variance  $N_0/2$.
Without loss of generality, we assume that $|h_1|\geq |h_2|$ and $R_1\geq R_2$ in this paper.

With respect to \eqref{eq:eq2}, let $x= h_1x_1+h_2x_2$ denote a superimposed symbol under channel coefficients $h_1$ and $h_2$ (i.e., the noise-free received signal). We define the set that collects all superimposed symbols under $h_1$ and $h_2$ as $\tilde{\chi}=\{h_1x_1+h_2x_2|x_1\in\chi_1, x_2\in\chi_2\}$. In addition, we denote $\mathbf{x}$ as  a vector of the superimposed symbols, $\mathbf{x}(i)\in\tilde{\chi}$, $i=1,2,...,L$.
We consider
block fading channels where $h_1$ and $h_2$ keep unchanged
during one block transmission and vary from one block to the
next. In general,  $|\tilde{\chi}|=q^2$.  We then define a  pair of coded messages $(c_1, c_2)$ as a joint message. The set of joint messages is given  by
\begin{align} \label{eq:jset}
\mathcal{C} =\{(c_1, c_2)| c_1, c_2\in \mathbb{F}_q\},
\end{align}
where  $|\mathcal{C}|= q^2$. Accordingly, each element in $\tilde{\chi}$ is associated with a joint message $(c_1, c_2)$ in $\mathcal{C}$.
 Thus, $\mathcal{C}\to \tilde{\chi}$  is a one-to-one mapping  under given fading channel coefficients. The receiver in GMAC aims to recover all user messages from the received signal $\mathbf{y}$.


 \subsection{Problem at a Glance
}

The capacity region of a two-user GMAC is given by the convex
hull of the set of rate pairs $(R_1;R_2)$ that satisfies \cite{bibmac}
\begin{align}   \label{eq:eq4}
 R_1 &< I(x_1; y|x_2)  \nonumber  \\
  R_2 &< I(x_2; y|x_1) \nonumber  \\
  R_1 + R_2 &< I(x_1, x_2; y),
\end{align}
for some joint distribution $(x_1;x_2)\sim p(x_1)p(x_2)$.  The SIC decoding   can achieve one of the two
corner points of the pentagonal region  \cite{bibmac}
\begin{align}   \label{eq:eq4}
 R_1 &< I(x_1; y),  \nonumber  \\
  R_2 &< I(x_2; y|x_1).
\end{align}
The remaining points on the dominant face  can be achieved by
time-sharing and rate-splitting \cite{bibmac}. Every rate pair in the interior
of the rate region can be achieved by simultaneous decoding.
Assuming that the messages are drawn uniformly at random, the average probability of error is defined as
\begin{align}
 P_e^{(n)} = P\{(\hat{u_1},\hat{u_2}) \ne (u_1,u_2)\}.
\end{align}
A rate pair $(R_1;R_2)$ is said to be achievable if there exists a
sequence of $(2^{nR_1}; 2^{nR_2}; n)$ codes such that $\lim\limits_{n\rightarrow \infty} P_e^{(n)} = 0$.

CFMA has been proposed to achieve some rate pairs of GMAC capacity without rate-splitting or time
sharing\cite{ITCFMA,CCFA} .
The scheme employs  nested lattice codes and a sequential decoding scheme. In the first step, the
receiver decodes a linear  combination of codewords $a_1\mathbf{u}_1 +a_2\mathbf{u}_2$
(modulo-lattice reduced) with non-zero integer coefficients
$a_1$ and $a_2$; in the second step, the receiver exploits this
linear combination as side information to decode one of the
codewords (either $\mathbf{u}_1$ or $\mathbf{u}_2$); finally, the receiver recovers
the other message from the outputs of  previous two steps. Furthermore, this decoding scheme achieves (a
part of) the dominant face of the capacity region for the GMAC \cite{CF15}.
 Later on,  rather than based on lattice codes, a generalization of CFMA  based on nested linear codes and
joint typicality encoding and decoding is presented \cite{CF115}.

In the linear-code-based GMAC, the linear-network coding $\phi$ mapping  from a superimposed vector $\mathbf{x}$ to a NC message vector c is defined as
\begin{align} \label{eq:eq45}
\mathbf{c}  &= \phi(\mathbf{x}) = \phi(h_1\mathbf{x}_1+h_2\mathbf{x}_2)=   a_1\mathbf{c}_1 + a_2\mathbf{c}_2 \nonumber \\
 &=  a_1\varphi_1^{-1}(\mathbf{x}_1)\oplus_q a_2\varphi_2^{-1}(\mathbf{x}_2),
\end{align}
 where $a_1,a_2\in \mathbb{F}_q$, $\oplus_q$ denotes the addition over $\mathbb{F}_q$, and    $\varphi_m^{-1}$ denotes the inverse  of $\varphi_m$, $m=1,2$.

%
%
%
 The following theorem   describes an achievable rate region
with CFMA.

{\it Theorem 1}  \cite{CF115}: A rate pair $(R_1, R_2)$ is achievable with
nested linear codes and under  CFMA decoding if for some non-zero coefficient vector $(a_1, a_2)\in \mathbb{F}^2_q$ and for some mapping $\varphi_1^{-1}$ and $\varphi_2^{-1}$, we have either
\begin{align}   \label{eq:8a}
 R_1 &< H(x_1) - \max\{H(c|y),H(x_1,x_2|y,c)\},  \nonumber  \\
  R_2 &<  H(x_1) - H(c|y),
\end{align}
or
\begin{align}   \label{eq:8b}
 R_1 &<   H(x_1) - H(c|y), \nonumber  \\
  R_2 &< H(x_2) - \max\{H(c|y),H(x_1,x_2|y,c)\},
\end{align}
where
\begin{align}   \label{eq:eq4}
 c&  = a_1c_1\oplus_q a_2c_2, \nonumber \\
 (c_1,c_2) &= (\varphi_1^{-1}(x_1),\varphi_2^{-1}(x_2)),
\end{align}
and
\begin{align}   \label{eq:eq441}
  H(c|y)=H(c)-I(c;y).
\end{align}

 \begin{figure}[!t]
\center
\includegraphics[width=3.3in,height=2.6in]{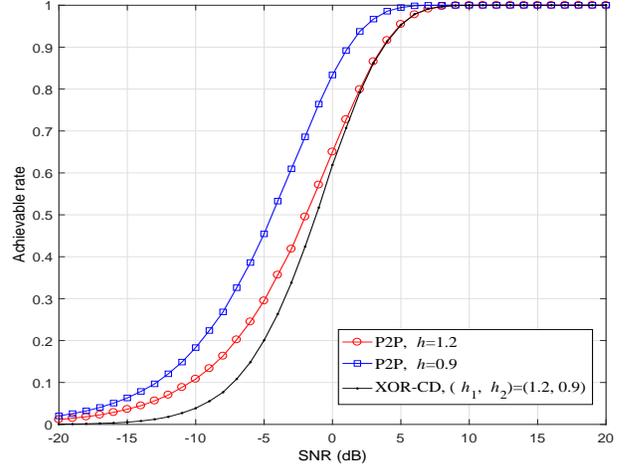}
\vspace{-0.3cm}
\caption{Achievable rates of XOR-CD with $(h_1, h_2) =(1.2, 0.9)$.
}
\label{fig:PNC11}
\end{figure}

 \begin{figure}[!t]
\center
\includegraphics[width=3.3in,height=2.6in]{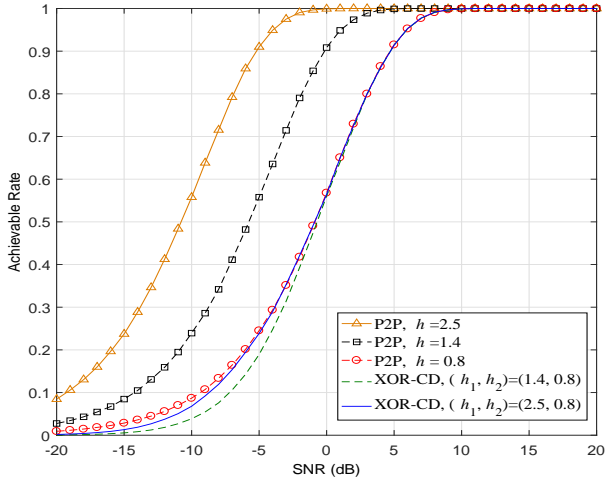}
\vspace{-0.3cm}
\caption{Achievable rates of XOR-CD with $(h_1, h_2) =(1.4, 0.8)$ and $(h_1, h_2) =(2.5, 0.8)$, respectively.
}
\label{fig:PNC22}
\end{figure}

 By Theorem 1, for some $(a_1, a_2)\in \mathbb{F}^2_q$, CFMA can achieve non-corner rate points   in \eqref{eq:8a} and \eqref{eq:8b}, which are distinct from corner points on the dominant face. 

It can be seen from \eqref{eq:8a}-\eqref{eq:eq441} that the rate $I(c; y)$ is vital in the achievable rate pairs. In the following, we illustrate  $I(c; y)$ in CFMA over different channel gains $h_1$ and $h_2$.
We will show that the achievable rate $I(c; y)$ is strictly limited by the worse channel of the two user channels, even though the other user channel is sufficiently good.
  We consider a  binary channel-coded two-user CFMA, i.e., $q$ = 2.  In this case, XOR-CD is used in CFMA to   decode the NC message vector $\mathbf{c}= \mathbf{c}_1\oplus \mathbf{c}_2$ with $a_1=a_2=1$ in \eqref{eq:eq45}. Fig. \ref{fig:PNC11} shows the achievable rate $I(c; y)$ of XOR-CD.
 The achievable rate of point-to-point (P2P) communication  with a channel gain $h$ is also plotted as an upper bound. We can see that XOR-CD with $(h_1, h_2) =(1.2, 0.9)$ has a rate loss
 as compared with P2P   with $h=0.9$ in the low-to-medium SNR regimes, and approaches the latter one in the high SNR regime. This rate loss is increased up to 3--5 dB as compared to P2P channel with $h=1.2$ over all SNR regimes.

We  enlarge the  difference of the two channel conditions and plot the results in Fig.  \ref{fig:PNC22}. First, it shows that XOR-CD  with $(h_1, h_2) =(1.4, 0.8)$ has a slight rate loss as compared with P2P   with $h=0.8$, and the rate loss is increased to  about 4--6 dB  with $h=1.4$. Second,   we note that  XOR-CD with $(h_1, h_2)=(1.4,0.8)$ performs close to XOR-CD even with a much larger $h_1$, e.g., $(h_1, h_2)=(2.5,0.8)$. From these results, one can deduce that  %
\begin{align}
 I(c;y)\leq\text{min}(C_1,C_2),
 \label{eq:eq33a}
\end{align}
where $C_1$ and $C_2$ denote the channel capacities of the users 1 and 2, respectively.
Thus, $I(c; y)$ has a rate loss compared to the capacity of the better channel, and the loss becomes larger as the channel difference becomes more significant.  Moreover,  the   performance loss of XOR-CD degrades the decoding performance of CFMA and NCMA, as the second step for these two schemes is to decode both user messages.  The simulation results in  \cite{nonp} have also shown that the uplink with the
poorest channel condition becomes the bottleneck of
NCMA, when   all users use the same modulation and code rate by ignoring
their individual channel gains.





\section{Proposed Channel Code and Decoder Designs}
This section develops a pair of encoder and decoder  for a two-user GMAC  to  employ different coding rates according to the channel conditions.
%
%
 First, we develop the channel codes   by the  RE and RC methods.
  Second, we propose a novel RDJD decoder to directly decode both   user messages rather than to decode the users' messages one by one, as in CFMA and NCMA.

%
%
%

%
%

\subsection{Proposed  Channel Code Designs}

We suppose that users 1 and 2 are the strong and weak users, respectively. We design two codes for the two users with the same codeword length but different code rates. We consider that user 1 adopts a code $\mathcal{C}_1$ of rate $R_1$ with  a parity-check matrix $\mathbf{H}_1$, while user 2 adopts a code $\mathcal{C}_2$ of  rate $R_2$ with a parity-check matrix $\mathbf{H}_2$. Note that $R_1\geq R_2$.
    Moreover, we design the codes that satisfy $\mathcal{C}_1\supseteq \mathcal{C}_2$.  That is,  the code $\mathcal{C}_2$  is a subcode of $\mathcal{C}_1$  given the same codeword length. In this way,
    the user messages can be decoded by a single decoder based on the low-rate parity-check matrix. To this end,   we propose  the RE and RC channel code design methods as follows.



%

\subsubsection{RE method}
 Given the parity-check matrix $\mathbf{H}_1$ of user 1, we construct the parity-check matrix $\mathbf{H}_2$ for $\mathcal{C}_2$ of user 2   with the lower rate $R_2$  by appending an additional matrix   $\mathbf{H}_e$ below in  $\mathbf{H}_1$, as shown in \eqref{eq:eqrc}.
  To reduce performance loss,  it is required that the newly added rows in $\mathbf{H}_e$ cannot introduce cycle-4   in  $\mathbf{H}_2$.
%

\begin{align} \label{eq:eqrc}
  \mathbf{H}_2 = \begin{array}{c@{\!\!\!}l}
  \left(\begin{array}[c]{ccccccccc}
       \multicolumn{9}{c}{\multirow{4}{*}{ \Large$\mathbf{H}_1$}}  \\\\\\\\
        \hline
    1 & 1 & 0 & 0 &\cdots &0 &1 &0 & 1 \\
    \vdots & \vdots & \vdots & \vdots& \vdots& \vdots& \vdots && \vdots \\
      \end{array}  \right)
&
 \begin{array}[c]{@{}l@{}l}\\\\\\\\
   \left. \begin{array}{c} \vphantom{0}  \\ \vphantom{0}
   \\ \end{array} \right\} & \text{$\mathbf{H}_e$} .
 \end{array}
\end{array}
\end{align}
%
%


\subsubsection{RC method}
 Given $\mathbf{H}_2$ of rate $R_2$, we first select some rows out of $\mathbf{H}_2$ to form a matrix $\mathbf{H}_r$; the matrix formed by the remaining rows in  $\mathbf{H}_2$ is denoted by $\mathbf{H}_d$, as shown in \eqref{eq:14a}. Second,
   every $\lambda$ rows in $\mathbf{H}_r$ are grouped to  generate a new row under some finite-field operations, and
    the resulting matrix $\mathbf{\overline   H}_r$ consisting of the new rows has fewer rows than $\mathbf{H}_r$. Finally, as shown in \eqref{eq:eq117}, we construct a new matrix $\mathbf{H}_1 = [\mathbf{H}_d;   \mathbf{\overline   H}_r]$ as    the parity-check matrix for user 1 with rate $R_1$. Obviously, $\mathbf{H}_1$   has fewer rows than $\mathbf{H}_2$. 

 Note that we can combine the rows in $\mathbf{H}_r$ in various ways, which may lead to different performance. For example, we can simply perform XOR
  on every $\lambda=2$ rows to generate $\mathbf{\overline   H}_r$. We require that the rows to be combined cannot have two or more ``1"s in the same column such that the variable-node degree distribution will remain unchanged.
\begin{subequations}
\begin{align} \tag{14a} \label{eq:14a}
  \mathbf{H}_2 = \begin{array}{c@{\!\!\!}l}
  \left(\begin{array}[c]{ccccccccc}
       \multicolumn{9}{c}{\multirow{4}{*}{ \Large$\mathbf{H}_d$}}  \\\\\\\\
        \hline
            1 & 0 & 0 & 1 &\cdots &0 &0 &0 & 1 \\
            0 & 1 & 0 & 1 &\cdots &1 &0 &0 & 1 \\
    \vdots & \vdots & \vdots & \vdots& \vdots& \vdots& \vdots && \vdots \\
      \end{array}  \right)
&
 \begin{array}[c]{@{}l@{}l}\\\\\\\\
   \left. \begin{array}{c} \vphantom{0}  \\ \vphantom{0}
   \\ \end{array} \right\} & \text{$\mathbf{H}_r$} \\
 \end{array}
\end{array}
\end{align}
\begin{align}\label{eq:eq117} \tag{14b}
  \mathbf{H}_1 = \begin{array}{c@{\!\!\!}l}
  \left(\begin{array}[c]{ccccccccc}
       \multicolumn{9}{c}{\multirow{4}{*}{ \Large$\mathbf{H}_d$}}  \\\\\\\\
        \hline
            1 & 1 & 0 & 0 &\cdots &1 &0 &0 & 0 \\
                \vdots & \vdots & \vdots & \vdots& \vdots& \vdots& \vdots && \vdots \\
      \end{array}  \right)
&
 \begin{array}[c]{@{}l@{}l}\\\\\\\\
   \left. \begin{array}{c} \vphantom{0}  \\ \vphantom{0}
   \\ \end{array} \right\} & \text{$\mathbf{\overline   H}_r$}.
 \end{array}
\end{array}
\end{align}
\end{subequations}
%
%
%
In particular, for  both   methods, we have
 \begin{align} \label{eq:eq17}
&\mathbf{c}_1 \mathbf{H}_1^T= {\mathbf 0} \ \ \  \text{for user 1}, \\ \nonumber
&\mathbf{c}_2 \mathbf{H}_2^T={\mathbf 0} \ \ \  \text{for user 2}.
\end{align}
%
%
%
%

From \eqref{eq:eqrc} and \eqref{eq:eq117},  there is a common parity-check matrix $\mathbf{H}_c$ between  $\mathbf{H}_1$  and   $\mathbf{H}_2$, and a residual matrix $\mathbf{H}_a$ in $\mathbf{H}_2$ besides $\mathbf{H}_c$. To be specific, $\mathbf{H}_c = \mathbf{H}_1$ and $\mathbf{H}_a=\mathbf{H}_e$ for RE, while  $\mathbf{H}_c = \mathbf{H}_d$ and $\mathbf{H}_a= \mathbf{\overline   H}_r$ for RC. It means that the codeword $\mathbf{c}_2$ checked by $\mathbf{H}_2$ is also a valid codeword checked by $\mathbf{H}_c$.  That is, $\mathbf{c}_2 \mathbf{H}_2^T=\mathbf{0}$ leads to $\mathbf{c}_2 \mathbf{H}_c^T=\mathbf{0}$, yielding
\begin{equation}\label{eq:eqqq17}
\left[
\begin{array}{c}
 \mathbf{c}_1  \\
 \mathbf{c}_2
\end{array}
\right ]\mathbf{H}_c^T= \mathbf{0}.
\end{equation}
Due to $\mathbf{c}    = \mathbf{ c}_1\oplus \mathbf{ c}_2$, we also have
\begin{equation}\label{eq:eqxor}
  \mathbf{c}     \mathbf{H}_c^T= \mathbf{0}.
\end{equation}
 \eqref{eq:eqxor}  means that we can decode $\mathbf{c}$ based on the high-rate parity matrix $\mathbf{H}_c$, as in CFMA and NCMA.

Moreover, from \eqref{eq:eqqq17}, we can employ a single decoder associated with $\mathbf{H}_c$ to decode both user messages. In addition to \eqref{eq:eqqq17},  $\mathbf{c}_2$ is also checked by $\mathbf{H}_a$, and we have
 \begin{align} \label{eq:eq118}
&\mathbf{c}_2 \mathbf{H}_a^T=\mathbf{0} \ \ \  \text{for user 2}.
\end{align}
Thus, we can employ an additional decoder associated with $\mathbf{H}_a$ to decode $\mathbf{c}_2$, which  can  provide independent information  to help decode both user messages.

Referring to the decoder  shown in  Fig. \ref{fig:CMFA1},  CFMA first decodes the NC message  vector $\mathbf{c}$ as $ \mathbf{\hat  c}$ according to \eqref{eq:eqxor} by XOR-CD and then decodes $\mathbf{c}_1$ as $\mathbf{\hat c}_1$ with the side information of $ \mathbf{\hat  c}$  according to \eqref{eq:eq17}.
  Note that XOR-CD in Step 1 uses the common high-rate parity-check matrix $\mathbf{H}_c$ to decode  the   vector $\mathbf{c}$ without involving the residual matrix $\mathbf{H}_a$  in low-rate $\mathbf{H}_2$. As a result, Step 1 cannot  fully exploit the decoding capability provided by  the low-rate code.
Thus, CFMA and NCMA with XOR-CD may not perform well when the two users have largely different coding rates.

 {\it Remark:} In the P2P communications, the better channel  allows a higher transmission rate to achieve reliable communication.
  In this context, in a two-user GMAC, it is desirable to allocate the strong user with a higher code rate and the weak user with a lower rate.   However, the rate allocation cannot work in XOR-CD, since  the achievable rate of XOR-CD is limited by the capacity of worse channel.   Thus, both NCMA and CFMA cannot fully exploit the different user channel conditions.


%



%

 \begin{figure}[!t]
\center
\includegraphics[width=3.3in,height=1.0in]{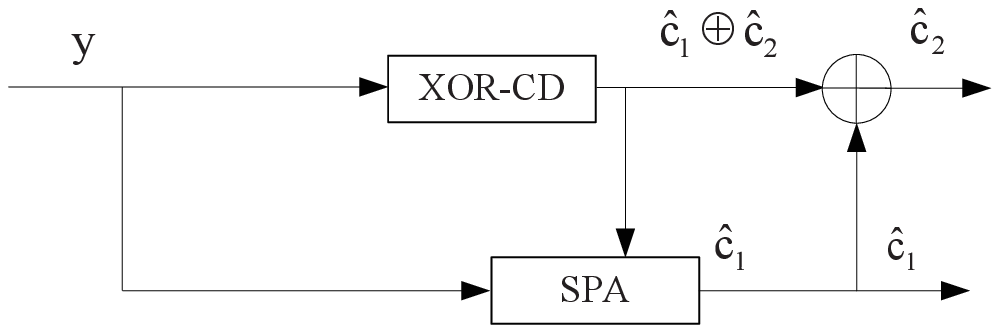}
\vspace{-0.3cm}
\caption{The decoding diagram of  CFMA.
}
\label{fig:CMFA1}
\end{figure}

\subsection{Proposed Decoder Design}

In contrast to CFMA that only relies on $\mathbf{H}_c$ to decode the NC message vector $\mathbf{c}    = \mathbf{ c}_1\oplus \mathbf{ c}_2$, our proposed decoder decodes the user messages using both $\mathbf{H}_c$ and the residual matrix $\mathbf{H}_a$ in $\mathbf{H}_2$. 


 To achieve this goal, we propose a novel RDJD decoding algorithm consisting of a JUD decoder  and a RUD decoder. Different from XOR-CD that decodes the NC message vector $\mathbf{c}$, JUD first estimates both user messages $\mathbf{c} _1$ and $ \mathbf{c} _2$ within $\mathbf{H}_c$ simultaneously. Second, extrinsic information about $\mathbf{c}_2$ from JUD is derived as  {\it a priori} information for  RUD, which can estimate $\mathbf{c} _2$ within $\mathbf{H}_a$ since $\mathbf{c} _2 \mathbf{H}_a^T = \mathbf{0}$. Then the extrinsic information of $\mathbf{c}_2$ is fed back from RUD to JUD to further  help decode $(\mathbf{c}_1, \mathbf{c}_2)$. As shown in Fig. \ref{fig:RDJD},   soft messages are iteratively passed between JUD and RUD. We summarize the decoding procedure of the RDJD as  follows.

%
\begin{framed}
\begin{enumerate}
\item Initiation: Set the maximum number of outer iterations to $G_{outer}$. Given the  received signal $\mathbf{y}$,  compute the initial channel messages $\mathbf{p}_{ch}$.

\item JUD: With  $\mathbf{p}_{ch}$  and {\it a priori} information of $\mathbf{c}_2$ from RUD,  JUD   decodes   $(\mathbf{c_1},\mathbf{c_2})$ with  $\mathbf{H}_c$ and outputs $\mathbf{\hat  c}_1$ and $\mathbf{\hat  c}_2$.
    If $\mathbf{\hat  c}_1 \mathbf{H}_1^T = \mathbf{0}$ and $\mathbf{\hat  c}_2 \mathbf{H}_2^T = \mathbf{0}$, or the number of iterations reaches  $G_{outer}$, stop decoding and output  $\mathbf{\hat  c}_1$ and $\mathbf{\hat  c}_2$ as the decoded codewords; otherwise, output extrinsic information of  $\mathbf{c}_1$ and $\mathbf{c}_2$ and go to step 3.
\item RUD: With  {\it a priori} information of  $(\mathbf{c} _1, \mathbf{c} _2)$, RUD estimates $\mathbf{c}_2$ by using the marginal messages of $\mathbf{c}_2$ with  $\mathbf{H}_a$. Pass the extrinsic information of $\mathbf{c}_2$ to JUD  and go to step 2.
\end{enumerate}
\end{framed}
%
%
 Next we elaborate JUD and RUD.
\subsubsection{JUD}
Define the joint   codeword as $\textbf{c}_{1,2}=[\textbf{c}_{1}; \textbf{c}_{2}]$ and  its $i$th element as $\textbf{c}_{1,2}(i)=[\textbf{c}_1(i),\textbf{c}_2(i)]$, which is a   vector containing two coded bits, $i=1,2,...,L$. Thus, $\textbf{c}_{1,2}(i)$ has 4 different combinations, $\textbf{c}_{1,2}(i)\in\mathcal{C}$.  Note that $\textbf{c}_{1,2}(i)$ can be decoded by the same channel decoder with $\mathbf{H}_c$   in \eqref{eq:eqqq17}.
 We use $\eta$ to denote the conversion function from  a binary vector to its corresponding decimal value, i.e., $o=\eta(\textbf{c}_{1,2}(i))=\eta([\textbf{c}_1(i),\textbf{c}_2(i)])$. Then, $o\in\Delta = \{0,1,2,3\}$. Conversely, we use $\eta^{-1}_m(o)$ to denote the $m$th bit in the bit vector converted from $o$, $m=1,2$  \footnote{For example, if $\textbf{c}_{1,2}(i)$ = [1\  0], then $o=\eta([1\  0]) = 2$. Conversely, $\eta^{-1}_2(2) = 0$.}.
 Moreover, we denote ${p}^o(i)=\text{Pr}\{\textbf{c}_{1,2}(i) = \eta^{-1}(o)\}$.


 Let a length-$4$ probability vector $\mathbf{p}(i)$ denote the soft message of $\mathbf{c}_{1,2}(i)$ passing within JUD. We perform sum-product algorithm (SPA) within $\mathbf{H}_c$ that consists of two sets of nodes, i.e., variable nodes $\mathbf{v}$ and check nodes $\mathbf{t}$. A variable node $\mathbf{v}(i)$ corresponds to  $\mathbf{c}_{1,2}(i)$.

$\bullet$ {\it Initial messages for JUD}. Recall that the superimposed symbol $x=h_1x_1+h_2x_2$ has four possible values,  $x\in \tilde{\chi}=\{h_1 +h_2, h_1 -h_2, -h_1 +h_2, - h_1 -h_2, \}$ and $\tilde{\chi}^j$  is the $j$th element in  $\tilde{\chi}$ for each received signal. We   introduce a length-4 channel probability vector  $\mathbf{p}_{ch}(i)$ for the $i$th variable node, where  the $j$th element
 $\mathbf{p}_{ch}^j(i)$ denotes the probability of $\mathbf{x}(i) = \tilde{\chi}^j$ and is computed by
 \begin{align}   \label{eq:eqp}
\mathbf{p}_{ch}^j(i)  &= \text{Pr}(\mathbf{y}(i)|\mathbf{x}(i)=\tilde{\chi}^j,h_1,h_2) \nonumber  \\
 & =\frac{1}{\beta\pi\sigma^2} \exp \bigg(-\frac{|\mathbf{y}(i)-\tilde{\chi}^j|^2}{2\sigma^2} \bigg),
\end{align}
where  $\beta$ is a normalization factor that ensures $\sum\limits_{j=0}^{3} \mathbf{p}_{ch}^j(i) =1$.
%
%
%
 %

In addition, JUD has {\it a priori} information which is the extrinsic soft information $\mathbf{e}^j(i)$ of user 2 coming from RUD, and $\mathbf{e}^j(i)$ denotes the probability of $\mathbf{c}_2(i) = j$, $j\in \{0,1\}$. Then,  the initial information for JUD can be computed by
\begin{align}   \label{eq:eq1137}
\mathbf{p}^o(i) =  \beta\mathbf{p}_{ch}^o(i)\mathbf{e}^j(i),  \ j = \eta^{-1}_2(o), \ o\in\Delta.
\end{align}
Note that $[\mathbf{e}^0(i)\ \mathbf{e}^1(i)] =[0.5,0.5]$ in the first outer iteration for all the bits.


 %
 %

$\bullet$ {\it Message update of variable nodes and check nodes}.
We use functions \text{VAR} and \text{CHK} to represent message updating at the variable nodes and check nodes, respectively.
 The explicit update rules of \text{VAR} and \text{CHK}  with degree three are given in Appendix A. The
messages from the variable nodes with degree greater than
three can be recursively calculated by
\begin{align}   \label{eq:eq5}
 \mathbf{v}_{\rightarrow k} =&{\text {VAR}}(\mathbf{p}(1), \ldots, \mathbf{p}(k-1), \mathbf{p}(k+1),\ldots,  \mathbf{p}(g) )   \nonumber \\
  =  &{\text {VAR}}(\mathbf{p}(g), {\text {VAR}}(\mathbf{p}(1), \ldots, \mathbf{p}(k-1), \nonumber \\
  & \mathbf{p}(k+1),\ldots,  \mathbf{p}(g-1))),
 \end{align}
where  $\mathbf{v}$ is a degree-$g$ variable node, $\mathbf{v}_{\rightarrow k}$ denotes the messages from $\mathbf{v}$ to its  $k$th connected check node, and $\mathbf{p}(k)$ denotes the input messages of $\mathbf{v}$, $k=1,2,...,g$.

Similarly, the messages from the check nodes with degree
 greater than three are calculated by
 \begin{align}   \label{eq:eq6}
\mathbf{t}_{\rightarrow k} =&{\text {CHK}}(\mathbf{q}(1), \ldots, \mathbf{q}(k-1), \mathbf{q}(k+1),\ldots,  \mathbf{q}(d) )  \nonumber  \\
 =  & {\text {CHK}}(\mathbf{q}(d), {\text {CHK}}(\mathbf{q}(1), \ldots,  \nonumber  \\
 & \mathbf{q}(k-1), \mathbf{q}(k+1),\ldots,  \mathbf{q}(d-1))),
 \end{align}
where  $\mathbf{t}$ is a degree-$d$ check node, $\mathbf{t}_{\rightarrow k}$ denotes the message from $\mathbf{t}$ to its $k$th connected  variable node, and $\mathbf{q}(k)$ denotes the input messages of $\mathbf{t}$, $k=1,2,...,d$.


%

$\bullet$ {\it Mapping of joint user messages}. After iterations, $\mathbf{v}(i)$ is estimated  as $o\in\Delta$.  The \textit{a-posteriori}  probability
 of $\mathbf{v}(i)$, denoted by  $\mathbf{w}(i) =[{w}^0(i),{w}^1(i), {w}^2(i),{w}^3(i)]$, is evaluated in \eqref{eq:eq6} by computing all
incoming messages from the check nodes as well as the  initial message. Finally, the joint  user messages are decoded as
\begin{align}   \label{eq:eq137}
 \mathbf{v}(i)   &= \eta^{-1}(o)=[\mathbf{c}_1(i), \mathbf{c}_2(i)], \nonumber \\
   \text{if}\  & \mathop{\text{argmax}}\limits_{\forall j\in \Delta} {w}^j(i) = o.  \end{align}


\subsubsection{RUD}
The information   $\bm{w}$ is then passed into RUD, which picks up the \textit{a priori}  probability of $\mathbf{c}_2$ .  Let $\mathbf{a}(i) =[\mathbf{a}^0(i),  \mathbf{a}^1(i)]$ with $\mathbf{a}^j(i)$ being the  probability of $\mathbf{c}_2(i)=j$, which can be computed by
\begin{align}   \label{eq:eq134}
\mathbf{a}^j(i) =  \sum\limits_{\eta^{-1}_2(o)=j, \atop o\in\Delta} {w}^o(i), i =1,2,...,L.
\end{align}
With  $\mathbf{a}(i)$ as the \textit{a priori}  probability,  $i=1,2,...,L$, we perform SPA decoder for the bit-wise  estimation of  $\mathbf{c}_2$ within $\mathbf{H}_a$. After decoding, we  generate {extrinsic} information $\mathbf{e}^0$ and $\mathbf{e}^1$ of $\mathbf{c}_2$ from the \textit{a posterior} probability by excluding $\mathbf{a}(i)$. The extrinsic information  then serves as  the \textit{a priori} information
to JUD in \eqref{eq:eq1137}.

From the above decoding procedure, we can regard RDJD as a
serial concatenated code, where soft information is serially updated between two constituent decoders JUD
and RUD.
Notably, the proposed rate-diverse GMAC outperforms SIC when two channel gains are even near-balanced and balanced.
This is due to the fact that two users adopt code rates of large difference to distinguish the user messages. RUD uses a  residual parity-check matrix  for one user, which helps JUD to decode message for the other user with the  common parity-check matrix. 

 \begin{figure}[!t]
\center
\includegraphics[width=3.3in,height=1.3in]{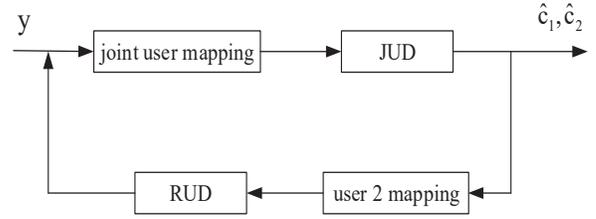}
\vspace{-0.3cm}
\caption{Decoding diagram of RDJD.
}
\label{fig:RDJD}
\end{figure}

\section{EXIT Analysis for Rate-diverse GMAC}
In this section, we   use an EXIT chart to illustrate the decoding behavior of RDJD by plotting the input/output relations of the constituent
 JUD and RUD decoders. This also provides an insight for rate allocation of  GMAC with diverse channel conditions.
%
%


%

Recall that the inputs to JUD are the channel information and the extrinsic information from RUD, while the inputs to RUD is the extrinsic information from JUD.
 Note that only the extrinsic information  for user 2 (i.e., the weak user) is iteratively updated between JUD and RUD.
Let $I_{{U_A}}$ denote the   {\it a priori} mutual information between the    coded bits and the input log-likelihood ratios (LLRs)   of JUD for user $2$;
and $I_{{U_E},m}$ denote the mutual information between  the coded bits and the output LLRs  of JUD for user $m$, $m=1,2$.
Similarly, let    $I_{{R_A}}$ and $I_{{R_E}}$ denote the {\it a priori} mutual information and extrinsic mutual information of  user $2$ from RUD, respectively.
The extrinsic output of RUD  is the {\it a priori} information for JUD  and vice versa.
We measure the  {\it a priori} mutual information  between
coded bit $X$ and the LLR $\xi$  by \cite{EXIT}
\begin{align}
I_{\rm A} = &\frac{1} {2} \sum_{\mu \in \{+1,-1\}} \int_{-\infty}^{\infty} f_{\rm A}
(\xi |X = \mu) \nonumber\\
&\times \log_2 \frac{2 f_{\rm A}(\xi|X = \mu)}
{f_{\rm A}(\xi|X = +1) + f_{\rm A}(\xi|X  = -1)}~{\rm d} \xi,
\label{eq:I-E}
\end{align}
where $f_{\rm A}$ is the conditional probability density function. By simulated observations, we can model    the {\it a priori} LLRs of the
constituent decoders   by an independent
Gaussian random variable with variance $\sigma^2_A$ and mean zero. Thus, we can compute $I_{{U_A}}$ and  $I_{R_A}$ using
  \begin{equation}
J(\sigma_{A}) = 1- \int_{-\infty}^{\infty}  \frac{\exp \left( -  \frac{(\xi -\sigma_{A}^2/2)^2 }{2 \sigma_{A}^2} \right)}{\sqrt{2\pi \sigma_{A}^2}}
\log_2 \left[ 1+\exp(-\xi) \right] {\rm d} \xi .
\label{eq:J_fuction}
\end{equation}
where $J(\sigma_{A})$ can be approximated by a polynomial \cite{EXIT}. Basically, an EXIT chart  in RDJD  includes two curves that characterize JUD and RUD, respectively.   Each curve represents a relation
between  the {\it a priori} information and  the extrinsic information. For each decoder, the mutual information  between the coded bit $X$ and the extrinsic LLR $\xi$  can be measured by
\begin{align}
I_{\rm E} = &\frac{1} {2} \sum_{\mu \in \{+1,-1\}} \int_{-\infty}^{\infty} f_{\rm E}
(\xi |X = \mu) \nonumber\\
&\times \log_2 \frac{2 f_{\rm E}(\xi|X = \mu)}
{f_{\rm E}(\xi|X = +1) + f_{\rm E}(\xi|X  = -1)}~{\rm d} \xi,
\label{eq:I-E}
\end{align}
where $f_{\rm E}(\xi|X = \mu)$ is the conditional PDF. 
For JUD,  $I_{{U_E},m}$ is defined as a function of $I_{{U_A}}$ and the channel information, i.e., SNR value, and is given by
\begin{equation}
I_{{U_E},m}   = T_U(I_{U_A}, SNR), \ m = 1, 2,
\label{eq:IUE}
\end{equation}
where $T_U$ represents JUD. Similarly, for RUD, we have
\begin{equation}
I_{{R_E}}   = T_R(I_{R_A}),
\label{eq:IRA}
\end{equation}
where $T_R$ represents RUD. Since the analytical treatments \eqref{eq:IUE} and \eqref{eq:IRA} for JUD and RUD are difficult to obtain, the distributions of $f_{\rm E}$ are most conveniently determined by Monte Carlo simulation (histogram measurements). Note that no Gaussian assumption is imposed on the distribution of the extrinsic outputs.

%
%
%
Given a received power $|h_1|^2+|h_2|^2=3.1$, we show the decoding behavior of the RDJD under a sum rate $R_1+R_2= 1.3$.
First, Fig. \ref{fig:LONG11} shows the transfer characteristics  $I_{{U_E},m}$, $m=1,2$,   from JUD
 at  a SNR  of  2.5  dB, where   the x-axis and y-axis denote the {\it a priori} input $I_{{U_A}}$ and the extrinsic
output $I_{{U_E},m}$, respectively. For all the rate pairs, we can see that $I_{{U_E},2}$ from JUD
are monotonically increasing in $I_{{U_A}}$. It is reasonable that the reliability of user 2 becomes larger
  because $I_{{U_E},2}   = T_U(I_{U_A}, SNR)$. In particular,   $I_{{U_E},1}$ from JUD also increases with $I_{U_A}$  although  $I_{U_A}$ is only related to  the reliability of user 2. It is because  JUD decodes the superimposed signal for the two user messages simultaneously, where the successful decoding of user 2 contributes to decoding convergence of user 1.

 \begin{figure}[!t]
\center
\includegraphics[width=3.0in,height=2.7in]{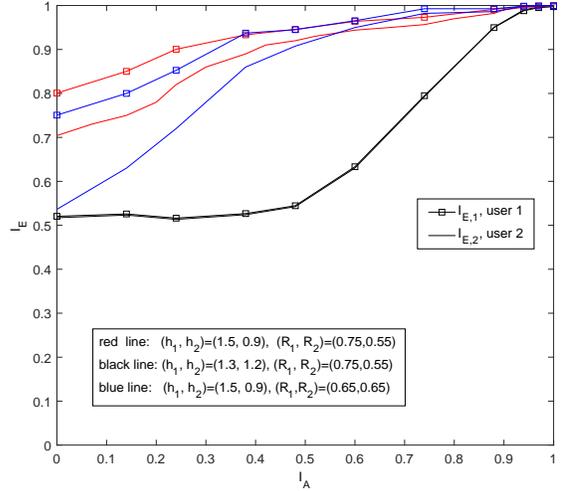}
\vspace{-0.3cm}
\caption{Extrinsic information transfer characteristics of soft in/soft out
 for JUD  at a SNR of 2.5 dB.
}
\label{fig:LONG11}
\end{figure}
 \begin{figure}[!t]
\center
\includegraphics[width=3.0in,height=2.7in]{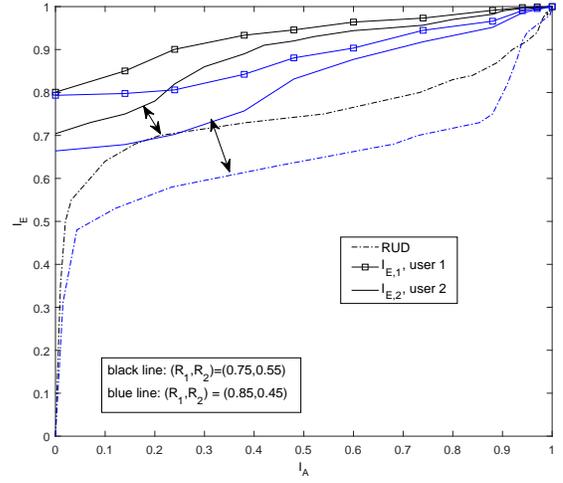}
\vspace{-0.3cm}
\caption{EXIT chart with transfer characteristics  of JUD and RUD at a SNR of 2.5 dB.
}
\label{fig:LONG199}
\end{figure}

Second,     Fig. \ref{fig:LONG11} shows the effect of different channel conditions on RDJD. The extrinsic mutual information of both users with $(h_1,h_2)=(1.5,0.9)$ is  much larger than  that with  $(h_1,h_2)=(1.3,1.2)$ given   $|h_1|^2+|h_2|^2=3.1$. Recall that JUD decodes four combinations, i.e., four superimposed symbols, of two user messages associated with $h_1$ and $h_2$. As the values of $h_1$ and $h_2$ are close, some of the superimposed symbols become close, which leads to the reduced Euclidean distance between the symbols.  In this case, the inter-user interference is equivalently severe and it is hard to distinguish  one user from the other user.


 Considering  $(h_1,h_2)=(1.5,0.9)$ and  $R_1+R_2= 1.3$, Fig. \ref{fig:LONG11} shows that $I_{{U_E},m}$ with rate pair $(R_1,R_2)=(0.75,0.55)$ is larger than that with    $(R_1,R_2)=(0.65,0.65)$, $m=1,2$. We also see that  $I_{{U_E},1}$ has a larger initial value than  $I_{{U_E},2}$ at the start point of $I_{U_A}=0$ due to the larger channel gain $|h_1|>|h_2|$. Accordingly, $I_{{U_E},2}$ has a gap from $I_{{U_E},1}$ during the iterations. However, as   $I_{U_A}$ for user 2 increases, $I_{{U_E},2}$ increases faster than $I_{{U_E},1}$ and the gap becomes narrower. In particular,  $I_{{U_E},2}$ is almost the same as $I_{{U_E},1}$ when $I_{U_A}$ becomes large enough, e.g., $I_{U_A}>0.6$.

Recall that JUD and RUD exchange extrinsic soft information iteratively.
  To account for the iterative nature, we also plot  the EXIT curves of JUD and RUD in Fig. \ref{fig:LONG199}.  The axes are swapped for transfer characteristics of the two
decoders. We define the region between the two expected EXIT curves, one for the JUD and one for RUD,   as the decoding tunnel.
%
%
%
   The figure plots  the EXIT curves of RDJD with   rate pairs $(R_1,R_2)=(0.75,0.55)$ and  $(R_1,R_2)=(0.85,0.45)$.  At a SNR of $2.5$ dB, the curves of JUD and RUD do not overlap for both cases. It means that RDJD can achieve a successful decoding with an arbitrarily small error probabilities for both cases.
Moreover, we can see that the case with $(R_1,R_2)=(0.85,0.45)$ has a larger decoding tunnel than that with $(R_1,R_2)=(0.75,0.55)$.
Since a larger decoding
tunnel implies a faster convergence rate and a lower decoding
threshold,   we deduce that the RDJD with $(R_1,R_2)=(0.85,0.45)$ converges faster and performs better than that with $(R_1,R_2)=(0.75,0.55)$.  Therefore, an optimized rate allocation is required to achieve the best performance.

\section{Simulation Result}
In this section, we provide numerical simulations of the
GMAC schemes  and compare with the theoretical rate
regions. For the theoretical rate regions, we evaluate theorem
1 with different discrete inputs. In our simulations, the total number of decoding iterations is set to 150.
For
fair comparison with iteratively decoded RDJD, the number of
iterations between JUD and RUD is 5, and
the number of channel decoding iterations is 30.  We
assume that a rate pair $(R_1, R_2)$ is achieved for a given power
and channel gains if the bit-error rate (BER) is below  $10^{-5}$
over 500 independent trials.  


\begin{figure}[!t]
\center
\includegraphics[width=3.3in,height=2.8in]{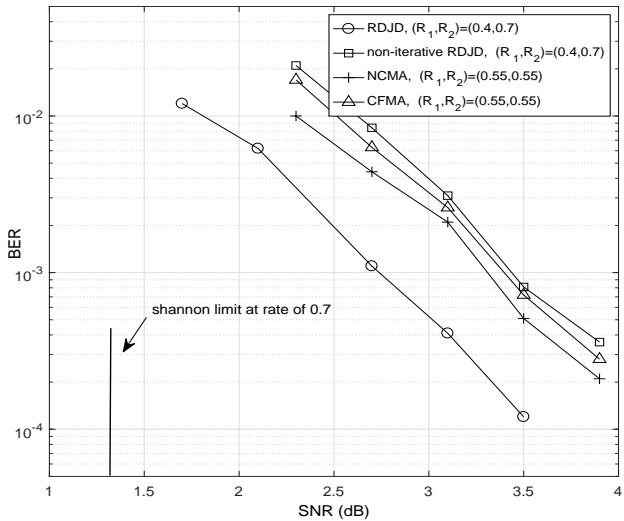}
\vspace{-0.3cm}
\caption{BER performance of GMAC schemes at $R_1+R_2=1.10$.
}
\label{fig:fig111}
\end{figure}

\begin{figure}[!t]
\center
\includegraphics[width=3.3in,height=2.8in]{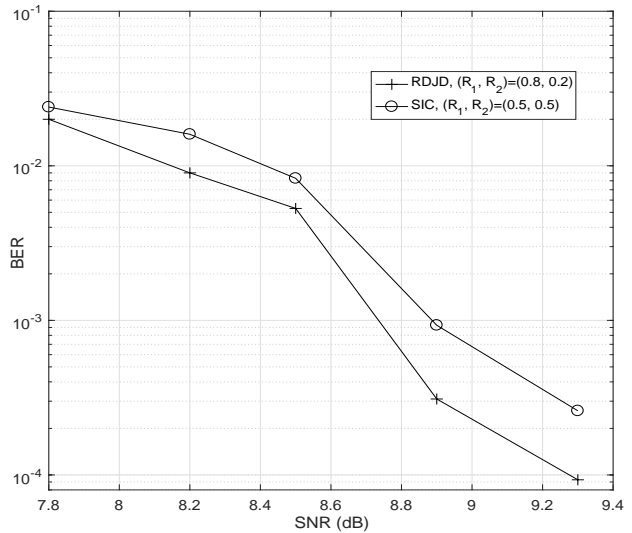}
\vspace{-0.3cm}
\caption{The performance of RDJD and SIC under near power-balanced GMAC.
}
\label{fig:blanc}
\end{figure}

\subsection{Average Performance of different GMAC schemes}

We compare  the average BER performance of the two users in different GMAC schemes.
Given a total transmission rate $R_1+R_2=1.1$, we first consider a GMAC channel of $(h_1, h_2)=(\sqrt{3},1.0)$.
For    $h_1=\sqrt{3}$, we employ  a (273,191) LDPC code with a parity-check matrix $\mathbf{H}_1$ of rate $R_1= 0.7$   \cite{SHULIN}.
For  $h_2=1.0$, we generate  a low-rate code $\mathbf{H}_2$  of $R_2=0.4$ by using our proposed RE method, which is achieved by appending 80 rows to $\mathbf{H}_1$. Each of the 80 rows has three ones    placed in randomly selected positions.
   Fig. \ref{fig:fig111}  shows that RDJD with iterative decoding has a large gain, about 0.8 dB, over   RDJD without iterations. It is consistent with the EXIT result that shows iterations between JUD and RUD  providing large gains.
  The figure also plots the BER of the conventional  NCMA and CFMA, where the  two users adopt the same parity-check matrix $\mathbf{H}$ with   $R_1=R_2=0.55$. The matrix $\mathbf{H}$ is generated by appending 40 rows to $\mathbf{H}_1$ by the RE method.
   We see that iterative RDJD has significant gains, 0.6 dB and 0.7 dB, over NCMA and CFMA at a BER of $3\times 10^{-4}$, respectively. Therefore, we can conclude that RUD   can help to distinguish and decode two user messages in JUD iteratively since it provides  independent  information for  the user 2. Moreover, we observe  a large gap between AWGN capacity and the decoding
performance of the existing GMAC schemes, and this gap can be greatly narrowed by the proposed RDJD.

Fig. \ref{fig:blanc} further compares the BER performance of the proposed scheme and the SIC scheme under near power-balanced
GMAC.  This scenarios is highly desirable in practice since  guaranteeing large received power differences between two users is not always possible \cite{nonp}.
  With $(h_1, h_2)=(1.0,0.95)$, two users in the proposed scheme is allocated with  different coding rates  $(R_1, R_2)=(0.8,0.2)$, while the users in SIC scheme adopt the same rate $(R_1,R_2)=(0.5,0.5)$.
 We can see from the figure that   RDJD outperforms   SIC by 0.4 dB at a BER of $3\times 10^{-4}$. It is because  the inter-user interference is large in  near power-balanced  scenarios. The effective SNR for decoding the user  messages is relatively small. The proposed scheme exploits  a large residual matrix in the low-rate code, which provides a larger decoding capability than the SIC in the case of strong  interference. In this way, we can decode one user message independently by RUD and use  it as side information to decode the other user message in JUD.

\subsection{Performance of RDJD with different rate pairs}
Next, we compare the BER  performance of two users with different rate pairs over the channels with  $(h_1, h_2)=(1.5, 0.9)$. Given  LDPC codes of block length 5000 bits and a sum rate of $R_1+R_2=1.3$,  we consider three pairs of $(R_1,R_2)$, namely  $(0.75,0.55)$, $(0.85,0.45)$, $(0.95,0.35)$. The high-rate code with $\mathbf{H}_1$ is the  LDPC code optimized for point-to-point AWGN communications. The additional parity-check matrix appended to  $\mathbf{H}_1$ to form the low-rate code of $\mathbf{H}_2$ is also optimized for  point-to-point communications.  Under iterative decoding, Fig. \ref{fig:fig222}  shows that the system with $(R_1,R_2)=(0.85,0.45)$ outperforms that with $(R_1,R_2)=(0.75,0.55)$ by 0.7 dB at a BER of $10^{-4}$. It is because the inferior channel $h_2=0.9$ requires a lower code rate to achieve successful decoding of user 2 in RUD.  Moreover, the system with $(R_1,R_2)=(0.85,0.45)$ yields a gain of 0.5 dB over that with  $(R_1,R_2)=(0.95,0.35)$  at a BER of $10^{-4}$. It is because that the common matrix $\mathbf{H}_1$ of $R_1=0.95$  provides less useful information than that with the lower rate $R_1=0.85$ for the first user. The first user using the high-rate $\mathbf{H}_1$  cannot be decoded successfully in JUD even with the reliable information of user 2 from RUD. The results suggest that there exists an optimal allocation of channel coding rates for the two users. In other words, given the channel gains and sum rate, we should consider proper rate allocation for the two users, which is associated with the common matrix and residual matrix  for JUD and RUD, respectively.


            Fig. \ref{fig:fig222} also shows that  RDJD with $(R_1,R_2)=(0.85,0.45)$ achieves a significant gain of 1.0 dB over  NCMA with the same rate $(R_1,R_2)=(0.65,0.65)$ at a BER of $5\times 10^{-4}$. Moreover, it is interesting to note that two users have much different decoding  performance  in  NCMA while they show  similar performance, within only 0.1 dB difference, for RDJD. It is because the two users have the same rate in NCMA but with different channel initial reliability caused by different channel gains. In RDJD, iterative information exchange between JUD for the two users and RUD for one user offers a performance balance between the two users. Moreover, with long codewords, Fig.~\ref{fig:figCC}   demonstrates that the proposed rate-diverse GMAC with $(R_1,R_2)=(0.85,0.45)$  can attain the non-corner  point of
the dominant face of capacity region.

\subsection{Performance of RDJD with low rate transmission}

We consider a low rate transmission with a sum rate $R_1+R_2=0.5$, $(h_1, h_2)=(1.4, 0.8)$, and a codeword length of 1024. NCMA   adopts a rate pair $(R_1, R_2)=(0.25,0.25)$ for the two users, and the channel codes used are the protograph LDPC codes \cite{lowpro}.  Fig. \ref{fig:fig3332} shows that   RDJD with $(R_1, R_2)=(0.34,0.16)$ outperforms  that with $(R_1, R_2)=(0.3,0.2)$ by 0.3 dB at a BER of $4\times 10^{-4}$.  For RDJD, the  parity-check matrices $\mathbf{H}_1$ of  $R_1$ and $\mathbf{H}_2$  of $R_2$ are generated from $\mathbf{H}$ by RC method and  RE method, respectively. We can see that the   RDJD with rate pair $(R_1, R_2)=(0.34,0.16)$ achieves a gain of 0.5 dB over  non-iterative RDJD.   The gains are less than that in the case  with $R_1+R_2=1.25$. It is because in the low rate transmission, both the parity-check matrices $\mathbf{H}_1$ and $\mathbf{H}_2$ are large. Thus, the additional decoding capability provided by $\mathbf{H}_a$ in $\mathbf{H}_2$ seems relatively low and has less effect  on  decoding both users' messages. However, the figure shows that the gain of RDJD over NCMA is still large, up to 0.8 dB, which is in part due to that JUD can yield a larger gain over XOR-CD at low rate than high rate.

\begin{figure}[!t]
\center
\includegraphics[width=3.3in,height=2.8in]{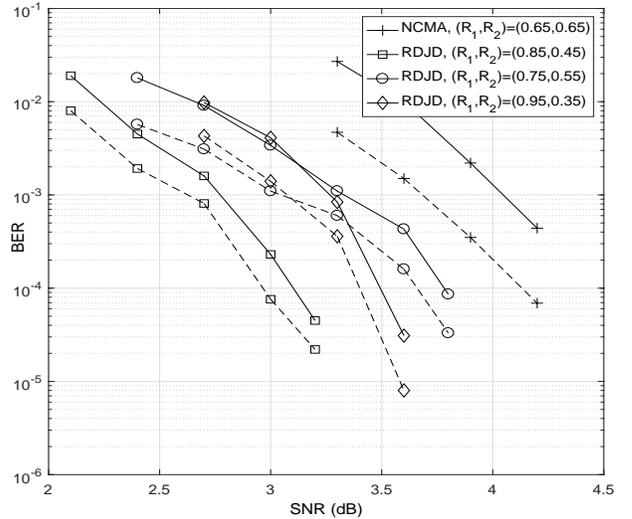}
\vspace{-0.3cm}
\caption{BER performance of RDJD  and NCMA at $R_1+R_2=1.3$.
}
\label{fig:fig222}
\end{figure}

\begin{figure}[!t]
\center
\includegraphics[width=2.8in,height=2.4in]{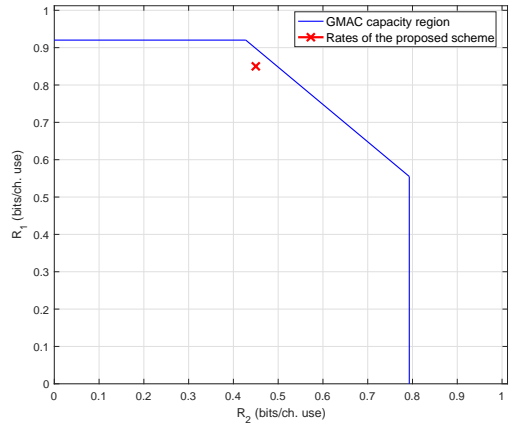}
\vspace{-0.3cm}
\caption{Capacity region and  rate pair of the proposed system.
}
\label{fig:figCC}
\end{figure}

\subsection{Performance of nonbinary channel-coded GMAC with 4PAM}
Fig. \ref{fig:nobinary} shows  the symbol error rate (SER) of GMAC schemes with 4PAM modulation and nonbinary LDPC coding over the integer ring $\mathbb{Z}_4$. Regular
LDPC codes with   codeword length of 600 symbols are used. The non-zero entries of the parity-check matrix are multiplicative invertible elements (non-zero divisors), i.e., \{1,3\}, randomly selected from $\mathbb{Z}_4$.
In this scenario,   JUD in RDJD is the generalized
nonbinary SPA (G-SPA) proposed in \cite{PNC1,PNC2}, and RUD is the conventional nonbinary SPA decoding.
Similarly, XOR-CD in CMFA and NCMA is the network-coding-based channel decoding (NC-CD) that performs symbol-by-symbol NC mapping prior to channel decoding.
Moreover, as in  \eqref{eq:eq45}, NC-CD attempts to
decode NC codeword with all possible coefficients, where $(a_1, a_2)$ are limited to non-zero divisor in  $\mathbb{Z}_4$, $(a_1, a_2) \in \{(1, 3),(1, 3),(3, 1),(3, 3)\}$.  Fig. \ref{fig:nobinary} plots the SER results for $(h_1, h_2) = (1.0, 0.8)$ and  $R_1+R_2=1.0$. We see that RDJD with $(R_1, R_2)=(0.6,0.4)$   has a
performance gain of 0.5 dB over NCMA in nonbinary channel-coded cases, while non-iterative RDJD with $(R_1, R_2)=(0.5,0.5)$ slightly outperforms NCMA by 0.1 dB.

\begin{figure}[!t]
\center
\includegraphics[width=3.3in,height=2.8in]{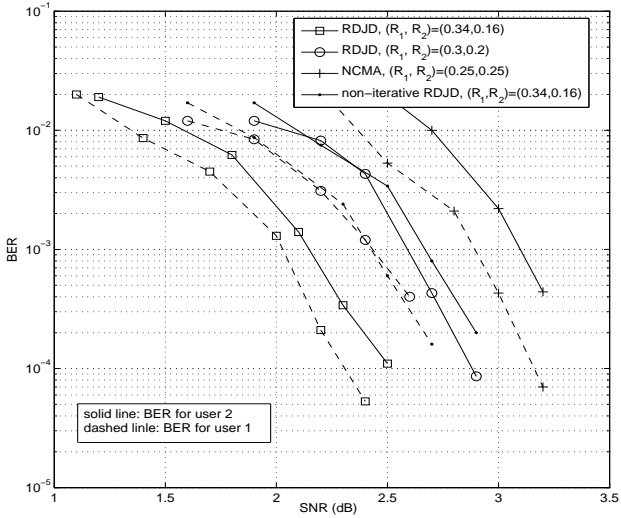}
\vspace{-0.3cm}
\caption{BER performance of RDJD  with different rate pairs at $R_1+R_2=0.5$.
}
\label{fig:fig3332}
\end{figure}

\begin{figure}[!t]
\center
\includegraphics[width=3.3in,height=2.8in]{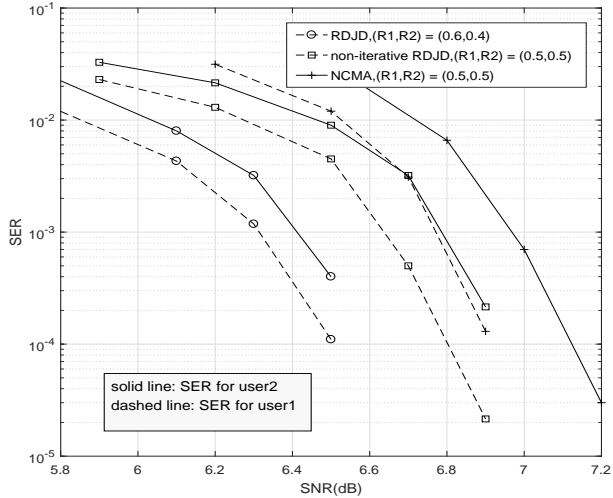}
\vspace{-0.3cm}
\caption{SER performance of nonbinary channel-coded GMAC schemes with 4PAM over $\mathbb{Z}_4$ at $R_1+R_2=1$.
}
\label{fig:nobinary}
\end{figure}


%
%
%
 \section{Conclusions}
%
%

In this paper, we   proposed a rate-diverse GMAC system that allows users to adopt different channel coding rate under different channel conditions.
In the GMAC system, we proposed two channel code design methods for the   two users such that the users with a good channel and an inferior channel employ  the channel codes of a high code rate  and a low code rate, respectively. Moreover, the two channel codes have some rows in their parity-check matrices identical.  We then proposed an iterative RDJD  as a serial concatenated decoder  that consists of JUD and RUD as two constituent codes. The JUD decodes both user messages within the common parity-check matrix, and RUD  provides independent information for the inferior  channel user protected by the residual parity-check matrix.
Moreover, we used the EXIT chart to show the decoding gain from the iterative decoding between JUR and RUD, and characterized the properties of different rate allocations for the two users.

To verify the EXIT analysis, we also  studied the BER performance of the proposed system when LDPC
channel codes were adopted. Our simulation
showed that the iterative RDJD achieved a large gain over the conventional NCMA and CFMA schemes  for both high rate and low rate transmissions. In particular, this gain could
be   more than 1 dB  at  relatively high rate cases.    Moreover, we found that there existed an optimal rate allocation to obtain the best decoding performance for the two users given a constant sum rate. It suggested that
the redundancy for JUD and RUD  should be optimized to achieve decoding convergence according to different channels in GMAC.

\section{Appendix A}

$\bullet$ {\it Output Message of Variable Nodes}.
Assume that the variable node $\mathbf{v}$ is connected to two check nodes and channel initial message. Given the two input messages $\mathbf{p}  =[p^0, p^1, p^2, p^3]$ of the initial message and $\mathbf{p}(1) =[p^0(1), p^1(1),p^2(1),p^3(1)]$ from the first check node, the outgoing probability message to the second check node,
denoted by $\mathbf{v}_{\rightarrow 2} = [{v}^0(2), {v}^1(2),{v}^2(2), {v}^3(2)]$, where ${v}^o(2)$ denotes the probability of $\mathbf{v}=o$, {is given by}
\begin{align}   \label{eq:eq99}
 &{v}^o(2) = \text{Pr}(\mathbf{v}=o|\mathbf{p},\mathbf{p}(1))  = \frac{\text{Pr}\{\mathbf{p},\mathbf{p}(1)|\mathbf{v}=o\}\text{Pr}\{\mathbf{v}=o\}}{\text{Pr}\{\mathbf{p},\mathbf{p}(1)\}} \nonumber\\
 &= \frac{\text{Pr}\{\mathbf{v}=o|\mathbf{p}\}\text{Pr}\{\mathbf{v}=o|\mathbf{p}(1)\}\text{Pr}\{\mathbf{p}\}\text{Pr}\{\mathbf{p}(1)\}}{\mathbf{v}\{\mathbf{v}=o\}\text{Pr}\{\mathbf{p},\mathbf{p}(1)\}}\\ \nonumber
 &= \beta p^op^o(1).
\end{align}
 Then, the output message is
\begin{align}   \label{eq:eq131}
 \mathbf{v}_{\rightarrow 2} = \text{VAR}(\mathbf{p}, \mathbf{p}(1)).
\end{align}

$\bullet$ {\it Output Message of Check Nodes}.
Suppose that a check node $\mathbf{t}$ is connected to three variable nodes. We use ${\mathbf{q}(1)}=[q^0(1),  q^1(1), q^2(1), q^3(1)]$ and  ${\mathbf{q}(2)}=[q^0(2),  q^1(2), q^2(2) ,q^3(2)]$ to denote the incoming messages from the variable nodes  $\textbf{v}(1)$ and $\textbf{v}(2)$, respectively. 
Let ${{\mathbf{q}}} = [q^0, q^1, q^2, q^3]$ denote the probability of
the coded bit $\mathbf{v}$ from $\mathbf{v}(1)$ and $\mathbf{v}(2)$, where  $q^o$ denotes the probability of $\mathbf{v}=o$.
We calculate message ${{\mathbf{q}}}$ based on ${\mathbf{q}(1)}$ and ${\textbf{q}(2)}$, with $q^o$ being
\begin{align}   \label{eq:eq9}
 {q}^o = \text{Pr}(\text{v}=d|\mathbf{q}(1),\mathbf{q}(2)) = \sum\limits_{\varphi(m,n)=o, \atop m,n\in\Delta} q^m(1)q^n(2),
\end{align}
where $\varphi(m,n)=o$, the bit-wise parity-check function of JUD, will be defined in \eqref{eq:eq11}, according to the LDPC encoding as below.

For   the $k$th user, the check node $\mathbf{t}$ is connected to three variable nodes $\mathbf{v}_k(1)$, $\mathbf{v}_k(2)$, and $\mathbf{v}_k(3)$, satisfying
\begin{align}   \label{eq:eq10}
 \textbf{c}_k(1)\oplus \textbf{c}_k(2) \oplus \textbf{c}_k(3) = 0,
\end{align}
where $k=1,2$. Thus, the encoding of JUD also satisfies the same requirement.
Suppose that three joint bit vectors corresponding to three variable nodes  are $\mathbf{c}_{1,2}(1) = [\mathbf{c}_1(1), \mathbf{c}_2(1)]$, $\mathbf{c}_{1,2}(2) =[\mathbf{c}_1(2), \mathbf{c}_2(2)]$, and $\mathbf{c}_{1,2}(3) =[\mathbf{c}_1(3), \mathbf{c}_2(3)]$, respectively. Moreover, $m$, $n$ and $o$ are used to represent the decimal values of the three vectors, i.e., $m=\eta(\mathbf{c}_{1,2}(1))$, $n=\eta(\mathbf{c}_{1,2}(2))$, and $o=\eta(\mathbf{c}_{1,2}(3))$.  Thus, with   respect to \eqref{eq:eq10}, given a value of $o$, we define $\varphi(m,n)=o$ in \eqref{eq:eq9} as
\begin{align}   \label{eq:eq11}
\eta^{-1}_k(m)\oplus \eta^{-1}_k(n) = \eta^{-1}_k(o),\ \ k = 1,2.
\end{align}
As defined before,   $\eta^{-1}_k(m)$ is the $k$th bit of the bit vector converted from a decimal value $m$.
%
%
%
Finally, the output message of $\mathbf{t}$ is
\begin{align}   \label{eq:eq134}
\mathbf{t}_{\rightarrow 3} =  \text{CHK}(\mathbf{q}(1), \mathbf{q}(2)).
\end{align}



\begin{thebibliography}{99}
%

\bibitem{han1}
T. S. Han and K. Kobayashi,  ``A new achievable rate region for the
interference channel,'' {\it IEEE Trans. Inf. Theory}, vol. 27, no. 1, pp. 49--60, 1981.

\bibitem{sic1} S. Verdu, {\it Multiuser Detection. Cambridge}, U.K.: Cambridge Univ. Press,
1998.





\bibitem{sic2}
  Z. Ding, Z. Yang, P. Fan, and H. Poor, ``On the performance of nonorthogonal
multiple access in 5G systems with randomly deployed
users,'' {\it IEEE Signal Process. Lett.}, vol. 21, no. 12, pp. 1501--1505, Dec.
2014.



\bibitem{han5}
B. Rimoldi and R. Urbanke,  ``A rate-splitting approach to the Gaussian
multiple-access channel,'' {\it  IEEE Trans. Inf. Theory}, vol. 42, no. 2, pp.
364--375, 1996.

\bibitem{han6}
A. J. Grant, B. Rimoldi, R. L. Urbanke, and P. A. Whiting,  ``Rate-splitting
multiple access for discrete memoryless channels,'' {\it IEEE Trans.
Inf. Theory}, vol. 47, no. 3, pp. 873--890, 2001.


{
\bibitem{added6}
J. Kelif, J. Gorce and A. Gati,  ``Performance and energy in green superposition coding wireless networks: an analytical model,'' {\it in GLOBECOM 2017}, Singapore, 2017, pp. 1-6.


\bibitem{SICNEW3} S. Vanka, S. Srinivasa, Z. Gong, P. Vizi, K. Stamatiou, and M. Haenggi, "Superposition coding strategies: design and experimental evaluation,''  {\it IEEE Trans. Wireless Commun.}, vol. 11, no. 7, pp. 2628--2639, 2012.

}


\bibitem{ccc3}
  S. M. R. Islam, N. Avazov, O. A. Dobre, and K.-S. Kwak, "Power-domain
non-orthogonal multiple access (NOMA) in 5G systems: potentials
and challenges," {\it IEEE Commun. Surveys Tuts.,} vol. 19, no. 2,
pp. 721--742, 2nd Quart., 2016.


\bibitem{dddd3}
A. Benjebbour, Y. Saito, Y. Kishiyama, A. Li, A. Harada, and
T. Nakamura,  ``Concept and practical considerations of non-orthogonal
multiple access (NOMA) for future radio access,'' {\it in Proc. ISPACS}, Nov. 2013,
pp. 770--774.


\bibitem{han2}
  R. Etkin, D. N. C. Tse, and H. Wang,  ``Gaussian interference channel
capacity to within one bit,'' {\it  IEEE Trans. Inf. Theory}, vol. 54, no. 12,
pp. 5534--5562, Dec. 2008.

\bibitem{han4}
B. Bandemer, A. El Gamal, and Y.-H. Kim,  ``Simultaneous nonunique
decoding is rate-optimal,'' {\it  in Proc. 50th Ann. Allerton Conf. Comm.
Control Comput.}, Monticello, IL, Oct. 2012.





\bibitem{han3}
H.-F. Chong, M. Motani, H. K. Garg, and H. El Gamal,  ``On the Han--
Kobayashi region for the interference channel,'' {\it  IEEE Trans. Inf. Theory},
vol. 54, no. 7, pp. 3188--3195, Jul. 2008.






%
%
%
%
%
%
%
%
%
%
%
%
%
%
%
%
%
%
%
%
%
%
%
%
%
%
%
%
%
%
%



\bibitem{bb6}
S. Chen, B. Ren, Q. Gao, S. Kang, S. Sun, and K. Niu,  ``Pattern division
multiple access (PDMA)--A novel non-orthogonal multiple access for 5G
radio networks,'' {\it IEEE Trans. Veh. Technol.}, vol. 66, no. 4, pp. 3185--3196,
Apr. 2017.

\bibitem{ccc12} Z. Yuan, G. Yu, W. Li, Y. Yuan, X. Wang, and J. Xu, ``Multi-user
shared access for Internet of Things,''  in {\it Proc. IEEE  VTC Spring}, May 2016, pp. 1--5.






\bibitem{SICNEW1} T. Cover, "Broadcast channels," {\it IEEE Trans. Inf. Theory.}, vol. 18, no. 1, pp. 2-14, Jan 1972.


\bibitem{SICNEW2}
N. I. Miridakis and D. D. Vergados, ``A survey on the successive interference cancellation performance for single-antenna and multiple-antenna OFDM systems,'' {\it IEEE Commun. Surveys Tutorials.}, vol. 15, no. 1, pp. 312--335, Feb. 2013.

\bibitem{RMAC0}
A. Roumy and D. Declercq, ``Characterization and optimization of LDPC
codes for the 2-user Gaussian multiple access channel,'' {\it EURASIP J.
Wireless Commun. Netw.}, vol. 2007, Art. no. 074890, 2007.

\bibitem{RMAC1}
A. Balatsoukas-Stimming and A. P. Liavas, ``Design of LDPC codes for the unequal power two-user Gaussian multiple access channel,'' {\it  IEEE  Wireless Commun. Lett.}, vol. 7, no. 5, pp. 868--871, Oct. 2018.

\bibitem{RMAC2}
S. Sharifi, A. K. Tanc and T. M. Duman, "LDPC code design for the two-user Gaussian multiple access channel,'' {\it  IEEE Trans. Wireless Commun.}, vol. 15, no. 4, pp. 2833-2844, April 2016.


\bibitem{RMAC3}
J. Du, L. Zhou, L. Yang, S. Peng, and J. Yuan, ``A new LDPC coded scheme for two-User Gaussian multiple access channels,'' {\it IEEE Commun. Lett.}, vol. 22, no. 1, pp. 21-24, 2018.



\bibitem{ccc6} L. Lu, L. You, and S. C. Liew, ``Network-coded multiple access,'' {\it IEEE
Trans. Mobile Comput.}, vol. 13, no. 12, pp. 2853--2869, Dec. 2014.

\bibitem{bib555}
   S. Zhang and S. C. Liew,
   ``Channel coding and decoding in a relay system operated with physical-layer network coding,'' {\it IEEE J. Sel. Areas Commun.}, vol. 27, no. 5, pp. 788--796, Oct. 2009.


\bibitem{you1}
L. You, S. C. Liew and L. Lu, ``Network-coded multiple access II: toward real-time operation with improved performance,'' {\it IEEE J. Sel. Areas Commun.}, vol. 33, no. 2, pp. 264--280, Feb. 2015.


\bibitem{you2}
H. Pan, L. Lu and S. C. Liew, ``Network-coded multiple access with high-order modulations,'' {\it  IEEE Trans. Veh. Technol.}, vol. 66, no. 11, pp. 9776-9792, Nov. 2017.

\bibitem{nonp}
H. Pan, L. Lu, and S. C. Liew, ``Practical power-balanced non-orthogonal multiple access," {\it IEEE J. Sel.
Areas Commun.}, vol. 35, no. 10, pp. 2312--2327, Oct. 2017.


\bibitem{uman}
H. Pan, S. C. Liew, J. Liang, Y. Shao and L. Lu, ``Network-coded multiple access on unmanned aerial vehicle,'' {\it IEEE J. Sel. Areas Commun.}, vol. 36, no. 9, pp. 2071--2086, Sept. 2018.




\bibitem{CCFA}
B. Nazer and M. Gastpar, ``Compute-and-forward: Harnessing interference through structured codes,'' {\it IEEE Trans. Inf. Theory}, vol. 57, no. 10,
pp. 6463--6486, Oct. 2011.

\bibitem{TAO}
T. Wang, S. C. Liew and L. Shi, ``Optimal rate-diverse wireless network coding,'' {\it IEEE Trans. Commun.}, vol. 65, no. 6, pp. 2411--2426, June 2017.


\bibitem{ITCFMA}
J. Zhu and M. Gastpar, ``Gaussian multiple access via compute-and-forward,'' {\it IEEE Trans. Inf. Theory}, vol. 63, pp. 2678--2695, May 2017.


\bibitem{CF15}
J. Zhu and M. Gastpar, ``On lattice codes for Gaussian interference channels,'' {\it 2015 IEEE International Symposium on Information Theory
(ISIT)}, Jun. 2015, pp. 2066-2070.

\bibitem{CF115}
 S. H. Lim, C. Feng, A. Pastore, B. Nazer, and M. Gastpar, ``A joint
typicality approach to algebraic network information theory,'' {\it ArXiv eprints,} Jun. 2016, preprint available at http://arxiv.org/abs/1606.09548.


\bibitem{CFMA}
E. Sula, J. Zhu, A. Pastore, S. H. Lim and M. Gastpar, ``Compute--forward multiple access (CFMA): practical implementations'', {\it IEEE Trans. Wireless Commun.}, vol. 67, no. 2, pp. 1133--1147, Feb. 2019.

\bibitem{EXIT}
S. ten Brink, ``Convergence behavior of iteratively decoded parallel concatenated codes,'' {\it IEEE Trans. Commun.}, vol. 49, no. 10, pp. 1727--1737, Oct. 2001.


\bibitem{SHULIN}
Y. Kou, S. Lin and M. P. C. Fossorier, ``Low-density parity-check codes based on finite geometries: a rediscovery and new results,'' {\it  Trans. Inf. Theory}, vol. 47, no. 7, pp. 2711--2736, Nov. 2001.


\bibitem{lowpro}

Y. Fang, G. Bi, Y. L. Guan and F. C. M. Lau, ``A survey on protograph LDPC codes and their applications,'' {\it IEEE Commun. Surveys Tuts.}, vol. 17, no. 4, pp. 1989--2016, Fourthquarter 2015.



\bibitem{bibmac}
T. M. Cover and J. A. Thomas, {\it Elements of Information Theory}, 2nd ed.
New York: Wiley, 2006.

\bibitem{PNC1}
P. Chen, L. Shi, S. C. Liew, Y. Fang and K. Cai, ``Channel decoding for nonbinary physical-layer network coding in two-way relay systems,'' {\it  IEEE Trans. Veh. Technol.}, vol. 68, no. 1, pp. 628--640, Jan. 2019.

\bibitem{PNC2}
P. Chen, S. C. Liew, and L. Shi, ``Bandwidth-efficient coded modulation
scheme  for physical-layer network coding with high-order modulations,'' {\it IEEE Trans. Commun.}, vol. 65, no. 1, pp. 147--160, Jan. 2017.
%

%


%






%
%
%
%
%


%











%




\end{thebibliography}
\end{document}